\documentclass[11pt,reqno,article,prd,superscriptaddress,onecolumn]{revtex4}
\usepackage{amsmath}%
\usepackage{amsfonts}%
\usepackage{amssymb}%
\usepackage{graphicx}
\usepackage{hyperref}

\begin{document}
\title{Effects of a Cut, Lorentz-Boosted sky on the Angular Power Spectrum}

\author{Thiago S. Pereira}
\affiliation{
Instituto de F\'isica Te\'orica, Universidade Estadual Paulista,
Rua Dr. Bento T. Ferraz, 271, 01140-070, S\~ao Paulo, SP, Brazil}
\affiliation{Departamento de F\'isica, Universidade Estadual de Londrina
Campus Universit\'ario, 86051-990, Londrina, Paran\'a, Brazil}

\author{Amanda Yoho}
\affiliation{
CERCA/ISO, Department of Physics, Case Western Reserve University,
10900 Euclid Avenue, Cleveland, OH 44106-7079, USA.}

\author{Maik Stuke}
\affiliation
{Falkult\"at f\"ur Physik, Universit\"at Bielefeld,
Postfach 100131, 33501 Bielefeld, Germany}

\author{Glenn D. Starkman}
\affiliation{
CERCA/ISO, Department of Physics, Case Western Reserve University,
10900 Euclid Avenue, Cleveland, OH 44106-7079, USA.}

%%%%%%%%%%%%%%%%%%%%%%%%%%%%%%%%%%%%%
\begin{abstract}
%%%%%%%%%%%%%%%%%%%%%%%%%%%%%%%%%%%%%
The largest fluctuation in the observed CMB temperature field is the dipole, its origin being 
usually attributed to the Doppler Effect -- the Earth's velocity with respect to the CMB rest frame. 
The lowest order boost correction to
temperature multipolar coefficients appears only as a second order correction in the
temperature power spectrum, $C_{\ell}$. Since $v/c\sim10^{-3}$, this effect can be safely
ignored when estimating cosmological parameters \cite{Challinor:2002zh,Kamionkowski:2002nd,Kosowsky:2010jm,Amendola:2010ty}.
However, by cutting
our galaxy from the CMB sky we induce large-angle anisotropies in the data. In this case,
the corrections to the cut-sky $C_{\ell}$s show up already at first order in the boost
parameter. In this paper we investigate this issue and argue that this effect might turn
out to be important when reconstructing the power spectrum from the cut-sky data.
\end{abstract}

\maketitle 

%%%%%%%%%%%%%%%%%%%%%%%%%%%%%%%%%%%%%
\section{Introduction}
%%%%%%%%%%%%%%%%%%%%%%%%%%%%%%%%%%%%%
In the last decade, cosmology has become a precision science, driven largely by 
measurements of the fluctuations in the Cosmic Microwave Background (CMB) temperature. 
These fluctuations -- nearly $10^{5}$ times smaller than the average temperature of the
universe -- are the primary window onto most  cosmological parameters
\cite{Larson:2010gs}. Because the fluctuations are so small, one must know the
contributions to the measured temperature field from foregrounds and other contaminants. 
A great deal of effort has been put in to characterizing foreground signals from dust,
synchrotron, and free-free emission \cite{Gold:2010fm}.  Even so, when inferring
cosmological parameters significant fractions of the sky are usually omitted 
(\lq\lq cut\rq\rq) from full-sky maps in order to minimize effects from non-primordial
sources \cite{Bennett:2003ca}. 

Surprisingly, one particular known systematic effect -- the distortion of the CMB
radiation due to our motion relative to the preferred cosmological frame -- has been given
comparatively little attention
\cite{Challinor:2002zh,Kamionkowski:2002nd,Kosowsky:2010jm,Amendola:2010ty}. 
It has become accepted practice to simply remove the dipole from a sky map before calculating the power spectrum
\cite{Hinshaw:2003ex}. This is based on the belief that the effect on $C_\ell$, the
$\ell$th element of the angular power spectrum of the temperature anisotropy, (the central
quantity in cosmological parameter estimation in the context of the canonical Lambda Cold
Dark Matter model, aka $\Lambda$CDM) is proportional to  $\beta^\ell$, where $\beta \sim
10^{-3}$ is our  speed relative to the cosmological frame (as determined by the magnitude
of that CMB dipole). In this context,  the dipole is indeed  the only multipole for which
the Doppler shift due to our motion is significant. Nevertheless, it has been 
shown \cite{Schwarz:2004gk} that the second order
(${\cal O}(\beta^2)$) Doppler effect noticeably alters the directions of the quadrupole multipole vectors,
which characterize the shape of the quadrupole, despite the fact that it contributes negligibly to $C_2$, the strength of the
quadrupole.

Meanwhile, several groups have independently examined (to order $\beta^2$) the effect of 
Lorentz boosting the CMB, and determined that  there are actually significant
contributions to higher
multipoles \cite{Challinor:2002zh,Kamionkowski:2002nd,Kosowsky:2010jm,Amendola:2010ty}. 
It has also separately been shown that simply masking a map induces correlations among
nearby multipoles \cite{Hivon:2001jp}. These two phenomena highlight a need for
characterizing the effect of masking a Lorentz-boosted temperature field, since the
process will undoubtedly mix higher multipoles of the true temperature field that have
boost corrections.

This paper is organized as follows: in Section \ref{boost_effect} we discuss the effect 
of boosting the multipole moments of the CMB, in Section \ref{masking_boost} we derive an
expression for the boosted pseudo power spectrum, in Section \ref{discussion} we show
preliminary numerical estimates for corrections to the power spectrum and discuss their
significance, as well as portions of the project that
will be explored in future papers.

%%%%%%%%%%%%%%%%%%%%%%%%%%%%%%%%%%%%%
\section{The effects of boosts on the multipole moments of the 
angular power spectrum of the CMB}\label{boost_effect}
%%%%%%%%%%%%%%%%%%%%%%%%%%%%%%%%%%%%%

If an observer in the rest frame of the CMB (denoted $S$) measures a  photon of frequency
$\nu$ arriving along a line of sight $\hat{\bf{n}}$, then an observer in another frame
$S^{\prime}$ that is moving with respect to the CMB at velocity $v\hat{\bf{v}}$
will measure the incoming photon to be arriving along a different line-of-sight
$\hat{\bf{n}}^{\prime}$ with a different frequency $\nu^\prime$. (Note that we will not
concern ourselves here with any ambiguities in determining S associated with the existence
of inhomogeneities,  in particular a cosmological dipole, on the assumption that that
dipole is ${\cal O}(10^{-5})$ smaller than the effects we will uncover.) The motion of
the observer in $S^{\prime}$ thus induces two effects: a Doppler shift  in the photon
frequency and an aberration  -- a shift in the direction from which the photon arrives.
These two effects can be seen explicitly in the relation between $\hat{\bf{n}}$ and
$\hat{\bf{n}}^{\prime}$
\begin{equation}\label{aberration}
\hat{\bf{n}}^{\prime}=\left(\frac{\cos\theta+\beta}{1+\beta\cos\theta}\right)\hat{\bf{v}}
+\frac{\hat{\bf{n}}-\cos\theta \hat{\bf{v}}}{\gamma (1+\beta \cos\theta )},
\end{equation}
where $\beta\equiv v/c$ and $\cos\theta\equiv\hat{\bf{n}}\cdot \hat{\bf{v}}$.
The change in observed frequency in $S^{\prime}$ is given by a simple Lorentz
transformation 
\begin{equation}\label{doppler}
\nu^{\prime}=\gamma \nu (1+\beta \cos\theta) \,,
\end{equation}
where $\gamma=(1-\beta^{2})^{-1/2}$ is the standard Lorentz factor. This angle $\theta$ is related to the angle
$\theta^{\prime}$, measured in the frame 
$S^{\prime}$, via 
\begin{equation}\label{angle}
\cos\theta' = { \frac{\cos\theta + \beta}{1+ \beta\cos\theta}} \, .
\end{equation}
Because of these two effects, we would like to boost the measured 
intensity, $I(\nu,\hat{\bf{n}})$ (the incident CMB power per unit area per unit frequency, per solid angle),
 and then relate the spherical harmonic
coefficients of the intensity to the traditional temperature fluctuation coefficients. First, we will need
the relation between the observed intensity in each frame
\cite{gravitationbook},
\begin{equation}
I^{\prime}(\nu,\hat{\bf{n}}^{\prime})=\gamma^{3}\left(1+\beta \cos\theta \right)^{3}
I(\nu, \hat{\bf{n}}).
\end{equation}
Expanding both sides in terms of spin-weighted spherical harmonics ($_{s}Y_{\ell m}$) and
using the fact that
$d\hat{\bf{n}}^{\prime}=\gamma^{-2}\left(1+\beta \cos\theta \right)^{-2}d\hat{\bf{n}}$
\cite{McKinley2}, we get the following expression for the boosted multipole moments,
$a_{\ell m}^{\prime}(\nu^{\prime})$, in terms of the rest-frame multipole moments 
$a_{\ell m}(\nu)$:
\begin{equation}\label{boost_alm}
a_{\ell_{1} m_{1}}^{\prime}(\nu^{\prime})=\sum_{\ell_{2},m_2}\int{d\hat{\bf{n}}
\gamma(1+\beta\cos\theta)a_{\ell_{2} m_{2}}(\nu)_{s}Y_{\ell_{2} m_{2}}
(\hat{\bf{n}})_{s}Y_{\ell_{1} m_{1}}^{\ast}(\hat{\bf{n}}^{\prime})}.
\end{equation}
We use the spin-weighted spherical harmonics to keep the expression completely general, so
that the effect of the boost on the measured CMB spectrum can be explored for temperature
fluctuations as well as polarization. Note that we have also implicitly chosen a frame where $\beta\bf{\hat{v}}$
is in the $\bf{\hat{z}}$ direction.

Expression (\ref{boost_alm}) can be expanded as a series in $\beta$ by means 
of Eqs. (\ref{aberration}) and (\ref{doppler}). The expression to order
$\beta^2$ can be found in the appendix. To order $\beta$ we find (consistent with
\cite{Challinor:2002zh}):
\begin{eqnarray}
\label{alm_freq_1}
a'_{\ell m}(\nu') & = & \left[1-\frac{\beta sm}{\ell(\ell+1)}
\left(2-\nu'\frac{d}{d\nu'}\right)\right]a_{\ell m}(\nu')\notag \\ 
&  &-\beta{}_{s}\xi_{(\ell+1)m}\left[(\ell-1)+\nu'\frac{d}{d\nu'}\right]
a_{(\ell+1)m}(\nu')-\beta{}_{s}\xi_{\ell m}
\left[-(\ell+2)+\nu'\frac{d}{d\nu'}\right]a_{(\ell-1)m}(\nu')
\end{eqnarray}
where
\[
{}_{s}\xi_{\ell
m}\equiv\sqrt{\frac{(\ell^{2}-m^{2})(\ell^{2}-s^{2})}{\ell^{2}(2\ell+1)(2\ell-1)}}\,.
\]
Here, the multipole moments are the frequency dependent intensity coefficients. The more
familiar temperature coefficients can be deduced from the coefficients above through the 
Stefan-Boltzmann law, which for fluctuations and a series expansion to $\mathcal{O}(\beta)$ reads:
\begin{equation}\label{stefan_boltz}
\frac{\delta T(\hat{\bf{n}})}{T_0}=\frac{1}{4}\frac{\delta I(\hat{\bf{n}})}{I_0}
\end{equation}
where
\begin{equation}\label{intensity_int}
I(\hat{\bf{n}})=\int_0^\infty I(\nu,\hat{\bf{n}})d\nu
\end{equation}
is the incident CMB power per unit area per solid angle and $T_0$ and $I_0$ are sky averages
(monopoles). Therefore, in order to get the temperature multipole moments we 
integrate Eq.(\ref{alm_freq_1}) over all frequencies (making use of the fact that the
$a'_{\ell m}(\nu')$ are Planck distributed) and divide the result by four times the
average intensity per solid angle \footnote{The monopole
$I_0$, being given by an integral over solid angle, has itself a Lorentz transformation.
However, we can always rescale its absolute value without introducing new
directionalities.}. To first order in $\beta$ we have for the temperature $a'_{\ell m}$ 
($s=0$):
\begin{equation}
\label{boost_alm_1storder}
a'_{\ell m}=a_{\ell m}+\beta\xi^+_{\ell m}a_{(\ell+1)m}+\beta\xi^-_{\ell m}a_{(\ell-1)m}
\end{equation}
where the conversion factor from temperature to intensity was absorbed into the multipolar
coefficients and
\begin{equation}
\xi^+_{\ell m}\equiv-(\ell-2)\,{}_0\xi_{(\ell+1)m}\,,\quad
\xi^-_{\ell m}\equiv(\ell+3)\,{}_0\xi_{\ell m}\,.
\end{equation}

We would now like to estimate the bias induced by the boost on the temperature power spectrum. Before
getting into the details of the calculation, it is first necessary to define two quantities. Throughout
this paper we will make the distinction between the theoretical power spectrum, given by
\begin{equation}
\langle a_{\ell m}a^{*}_{\ell^{\prime} m^{\prime}}\rangle=
	\delta_{\ell,\ell^{\prime}}\ \delta_{m,m^{\prime}}\ C_{\ell},
\end{equation}
and the measured power spectrum (also referred to as the power spectrum estimator), given
by
\begin{equation}
\mathcal{C}_{\ell}\equiv\frac{1}{2\ell +1}\sum_{m}\lvert a_{\ell m}\rvert^{2}.
\end{equation}
To estimate this bias we need to go second order 
in the expansion (\ref{boost_alm_1storder}) since the $C_\ell$s are quadratic in $a_{\ell m}$
(see the appendix for details). The key point here is that, on the assumption of statistical isotropy of the 
unboosted $a_{\ell m}$, the smallest $\beta$ correction to the boosted power spectrum is given 
by \cite{Challinor:2002zh}:
\begin{equation}
\langle {\mathcal{C}}'_\ell\rangle \approx C_\ell(1+4\beta^2+\mathcal{O}(\beta^3))\,.
\end{equation}
Note that the main effect is a rescaling of the spectrum by an overall amplitude $1+4\beta^2$.
Moreover, since $\beta\sim10^{-3}$, the boosted power spectrum is essentially unbiased. We will 
now show that this conclusion may not be true in the case where the boosted spectrum 
is reconstructed from a cut-sky.
%%%%%%%%%%%%%%%%%%%%%%%%%%%%%%%%%%%%%
\section{Masking effects on the boosted angular power spectrum}
\label{masking_boost}
%%%%%%%%%%%%%%%%%%%%%%%%%%%%%%%%%%%%%
When analyzing CMB temperature maps, it is common practice to mask regions of the sky 
that are believed to be contaminated. The region masked most often is the galaxy, where
the observed temperature signal is known not to come from the surface of last scattering.
In this case, we have a new expression for the measured temperature fluctuations:
\begin{equation}\label{pseudotemp}
\Delta\tilde{T}(\textbf{n})=\sum_{\ell>0}\sum_{m=-\ell}^{\ell}a_{\ell m}W(\textbf{n})
Y_{\ell m}(\textbf{n}),
\end{equation}
where $W(\textbf{n})$ is a window function described by the mask. One can then 
decompose the left hand side of Eq.(\ref{pseudotemp}) and solve for $\tilde{a}_{\ell m}$
in terms of $a_{\ell m}$. We then get the linear relation
\begin{equation}
\tilde{a}_{\ell_{1}m_{1}}=\sum_{\ell_{2},m_{2}}a_{\ell_{2} m_{2}}K_{\ell_{1} m_{1}
\ell_{2}m_{2}}.
\end{equation}
Here $K_{\ell_{1} m_{1}\ell_{2}m_{2}}$ is a general kernel that contains all of the 
information about the window function,
$W(\textbf{n})=\sum_{\ell_{1}m_{1}}w_{\ell_{1}m_{1}}Y_{\ell_{1}m_{1}}(\textbf{n})$.
It is given explicitly by 
\begin{eqnarray}
\label{kernel}
K_{\ell_{1} m_{1}\ell_{2}m_{2}}\equiv \sum_{\ell_{3},m_{3}}w_{\ell_{3}m_{3}}(-1)^{m_{2}}
\left[ \frac{(2\ell_{1}+1)(2\ell_{2}+1)(2\ell_{3}+1)}{4\pi} \right]^{\frac{1}{2}} \notag \\ \times 
\left(\begin{array}{ccc}
\ell_{1} & \ell_{2} & \ell_{3} \\
0 & 0 & 0 \end{array}\right)
\left( \begin{array}{ccc}
\ell_{1} & \ell_{2} & \ell_{3} \\
m_{1} & -m_{2} & m_{3} \end{array}\right),
\end{eqnarray}
where the $3\times2$ matrices are the Wigner 3-j symbols. This can be used to 
find the pseudo power spectrum estimator:
\begin{equation}
\mathcal{\tilde{C}}_{\ell}=\frac{1}{2\ell +1}\sum_{m=-\ell}^{\ell} \lvert 
\tilde{a}_{\ell m}\rvert^{2}.
\end{equation}
These $\mathcal{\tilde{C}}_{\ell}$s are not unbiased estimators for the true $C_\ell$s, 
but assuming statistical isotropy their 
expectation values are related by a real and symmetric mode coupling matrix,
\begin{equation}
\label{pseudo_Cls}
\langle\tilde{\mathcal{C}}_{\ell_{1}}\rangle=\sum_{\ell_{2}}
\mathcal{M}_{\ell_{1}\ell_{2}}\langle\mathcal{C}_{\ell_{2}}\rangle =\sum_{\ell_{2}}
\mathcal{M}_{\ell_{1}\ell_{2}}C_{\ell_{2}},
\end{equation}
where 
\begin{equation}
\label{matrixM}
\mathcal{M}_{\ell_{1}\ell_{2}}=\frac{2\ell_{2}+1}{4\pi}\sum_{\ell_{3}}(2\ell_{3}
+1)\mathcal{W}_{\ell_{3}} \left( \begin{matrix} \ell_{1} & \ell_{2} & \ell_{3} \\
0 & 0 & 0 \end{matrix}\right)^{2}
\end{equation}
and $\mathcal{W}_{\ell}=1/(2\ell+1)\sum_m|w_{\ell m}|^2$ is the power spectrum of the
window function \cite{Hivon:2001jp}.

We can therefore obtain an estimator for $C_{\ell}$ by approximately re-writing the relation with 
$\langle \mathcal{\tilde{C}}_{\ell}\rangle$ replaced by its observed value on the sky 
$\mathcal{\tilde{C}}_{\ell_{1}}^{obs}\simeq \sum_{\ell_{2}}\mathcal{M}_{\ell_{1}\ell_{2}}\mathcal{C}_{\ell_{2}}$.
We approximately invert Eq.(\ref{pseudo_Cls}) to
obtain the true $C_\ell$s from the pseudo ones. However, this prescription is
complicated by Lorentz boosts. Since the masking procedure is
unavoidably performed in the boosted frame, we no longer get a simple mode-coupling
relation as in Eq.(\ref{pseudo_Cls}). Instead, we will now have a more general expression:
\begin{eqnarray}\label{pseudo_aniso}
\langle\mathcal{\tilde{C}}_{\ell_{1}}^{\prime}\rangle&=&\frac{1}{2\ell_1+1}
\sum_{m_1=-\ell_1}^{\ell_1}|\tilde{a}'_{\ell_1 m_1}|^2 \notag \\
&=&\frac{1}{2\ell_{1}+1}\sum_{m_{1}=-\ell_{1}}^{\ell_{1}}
\sum_{\ell_{2}m_{2}}\sum_{\ell_{3}m_{3}}\langle a_{\ell_{2} m_{2}}^{\prime}
a_{\ell_{3} m_{3}}^{\ast\prime}\rangle K_{\ell_{1} m_{1}\ell_{2}m_{2}}
K_{\ell_{1}m_{1}\ell_{3}m_{3}}^\ast,
\end{eqnarray}
where the multipolar coefficients are given by Eq.(\ref{boost_alm_1storder}). Note that
the kernels in this expression couple all the elements of the non-diagonal covariance matrix
$\langle a_{\ell_1m_1}^{\prime}a_{\ell_2m_2}^{\ast\prime}\rangle$. As a consequence,
linear $\beta$ terms which were absent in Eq.(\ref{boost_alm_1storder}) will now contribute 
to the window function multipolar coefficients.

Going back to expression Eq.(\ref{pseudo_aniso}), note that to first order in $\beta$
there will only be coupling between $\ell_{2}$ and $\ell_{3}$ satisfying $|\ell_{2}-1|\leq\ell_{3}\leq\ell_{2}+1$. 
Therefore this expression can be re-written as
\begin{equation}
\langle \mathcal{\tilde{C}}_{\ell_{1}}^{\prime}\rangle =\frac{1}{2\ell_{1} +1}\sum_{m_{1}=
-\ell_{1}}^{\ell_{1}}\sum_{\ell_{2},m_{2}}\sum_{m_{3},\ell_{3}=\ell_{2}-1}^{\ell_{2}+1}
\langle
a_{\ell_{2}m_{2}}^{\prime}a_{\ell_{3} m_{3}}^{\ast\prime}\rangle
K_{\ell_{1}m_{1}\ell_{2}m_{2}}K_{\ell_{1} m_{1}\ell_{3}m_{3}}^\ast.
\end{equation}
When writing out each term in the sum over $\ell_{3}$ and plugging in the expression for 
$a_{\ell_{2} m_{2}}^{\prime}$ (see Eq.(\ref{boost_alm_1storder})), we arrive at the following 
relation between the boosted pseudo-$\mathcal{C}_{\ell}$'s and the expectation values of the CMB rest frame, $C_{\ell}$'s:
\begin{equation}
\langle\mathcal{\tilde{C}}'_{\ell_{1}}\rangle=\sum_{\ell_{2}}
\left(\mathcal{M}_{\ell_{1}\ell_{2}}C_{\ell_{2}}+\mathcal{N}_{\ell_1\ell_2}C_{\ell_2}
+\mathcal{P}_{\ell_1\ell_2}C_{\ell_2+1}+\mathcal{Q}_{\ell_1\ell_2}C_{\ell_2-1}
\right)
\end{equation}
where
\begin{eqnarray}
\label{matrixN}
\mathcal{N}_{\ell_{1}\ell_{2}} & = &
\frac{\beta}{2\ell_{1}+1}\sum_{m_{1},m_{2}}K_{\ell_{1}m_{1}\ell_{2}m_{2}}\left(K_{\ell_{1}
m_{1}(\ell_{2}+1)m_{2}}^{*}\xi_{(\ell_{2}+1)m}^{-}+K_{\ell_{1}m_{1}(\ell_{2}-1)m_{2}}^{*}
\xi_{(\ell_{2}-1)m}^{+}\right)\,,\\
\mathcal{P}_{\ell_{1}\ell_{2}} & = &
\frac{\beta}{2\ell_{1}+1}\sum_{m_{1},m_{2}}K_{\ell_{1}m_{1}\ell_{2}m_{2}}K_{\ell_{1}m_{1}
(\ell_{2}+1)m_{2}}^{*}\xi_{\ell_{2}m_{2}}^{+}\,,\\
\label{matrixQ}
\mathcal{Q}_{\ell_{1}\ell_{2}} & = &
\frac{\beta}{2\ell_{1}+1}\sum_{m_{1},m_{2}}K_{\ell_{1}m_{1}\ell_{2}m_{2}}K_{\ell_{1}m_{1}
(\ell_{2}-1)m_{2}}^{*}\xi_{\ell_{2}m_{2}}^{-}\,.
\end{eqnarray}
Ignoring "fence-post" terms and re-indexing the sum, we can re-write
Eq.(\ref{boostedpseudo_1storder}) as
\begin{equation}
\label{boostedpseudo_1storder}
\langle\tilde{\mathcal{C}}'_{\ell_{1}}\rangle\approx\sum_{\ell_{2}}
\left(\mathcal{M}_{\ell_{1}\ell_{2}}+\mathcal{N}_{\ell_1\ell_2}
+\mathcal{P}_{\ell_1\ell_2-1}+\mathcal{Q}_{\ell_1\ell_2+1}\right)C_{\ell_2}
\end{equation}
Clearly, the boosted pseudo power spectra in Eq.(\ref{boostedpseudo_1storder}) are far
from being unbiased estimators of the $C_\ell$s. 
If we neglect the boost effects on the $\mathcal{C}_{\ell}$s and only multiply by $\mathcal{M}^{-1}$ to solve for the values of
the $C_{\ell}$s, then we will obtain an unbiased estimate since $\mathcal{N}$, $\mathcal{P}$, and $\mathcal{Q}$ are $\beta$-dependent.  
Furthermore, the bias in Eq.(\ref{boostedpseudo_1storder})
produces more than just an overall multiplicative constant, since the matrices $\mathcal{M}$, $\mathcal{N}$, 
$\mathcal{P}$ and $\mathcal{Q}$ are not in general peaked at $\ell_2\approx\ell_1$. The correct reconstruction of the 
true $C_\ell$s depends on the careful inversion of the sum of the four matrices above. 
Before carrying a complete numerical analysis, we must analyze the type of couplings that the matrices $\mathcal{N}$, 
$\mathcal{P}$ and $\mathcal{Q}$ induce on the pseudo-power spectrum. This is the subject of the next section.
\begin{figure}[ht]
\includegraphics[scale=.48]{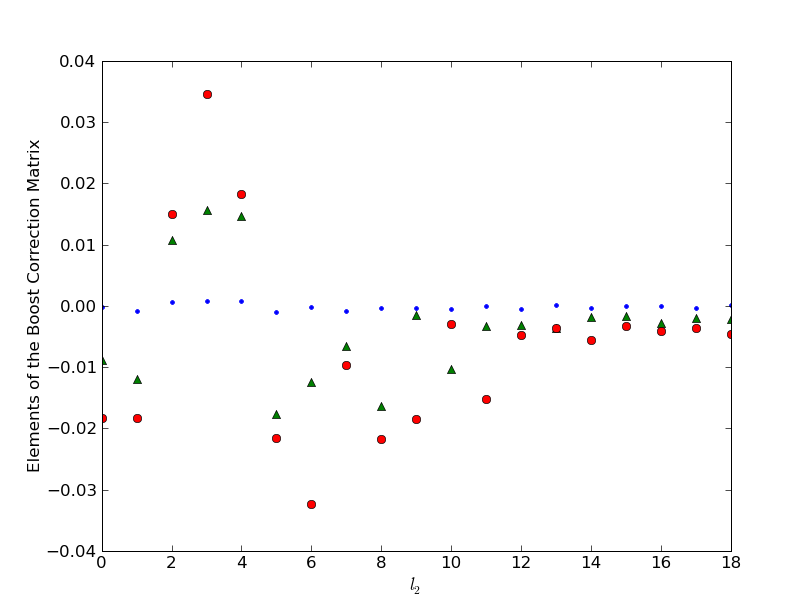}
\includegraphics[scale=.48]{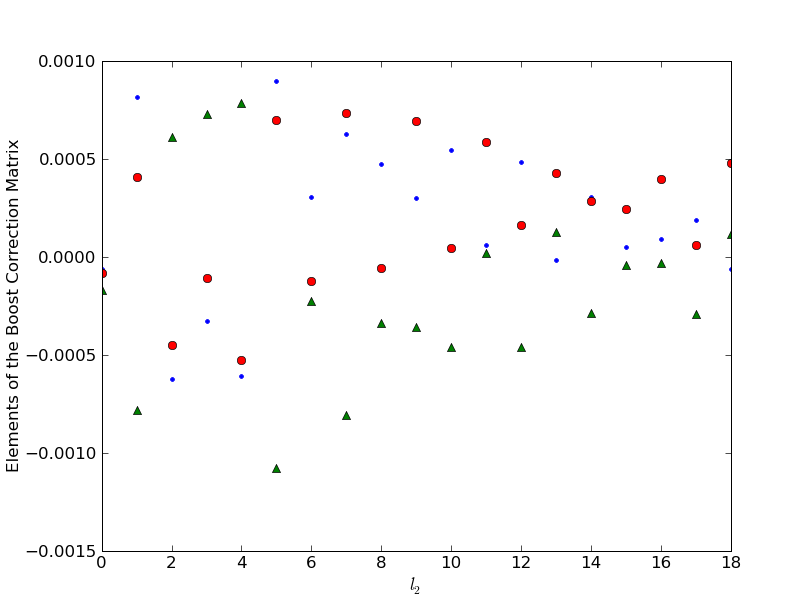}

\caption{\label{fig1}The top figure shows the effect of moving the mask north by 5 degrees with respect
to the galactic equator. Note that there are significant changes in contribution to the correction
matrix elements as the mask is moved to 76-105 degrees (blue dots), 86-115 degrees (green triangles), 
and 96-125 degrees (red circles). The bottom figure shows the effect of varying mask size while keeping
it only 1 degree north-south asymmetric for 76-105 (green triangles), 79-100 (blue dots),  and 84-95 (red circles)
degree masks. Here we notice slight changes in the matrix elements when the masks decrease in size, however the effect is
very small when compared to the effect of moving the mask off-center. All of these values were generated with $\ell_1=5$.}
\end{figure}

%%%%%%%%%%%%%%%%%%%%%%%%%%%%%%%%%%%%%
\subsection{Example: Equatorial Strip}
%%%%%%%%%%%%%%%%%%%%%%%%%%%%%%%%%%%%%

Before we calculate these couplings explicitly, we should mention two important
points. First, in this section we will examine only the special case where 
both the boost and the normal to the mask are  in the $z$-direction. 
This is of course unlikely to be correct.  It is also likely to underestimate the magnitude
of the effects we are examining.  A thorough analysis 
that also includes the angle between the boost and the galactic plane as a free parameter 
is being  carried out and will be reported on in a future work. 
Our point here is to note that $\beta$ corrections exist regardless of this
direction, and might in turn affect the reconstruction of the true $C_\ell$s if the boost
is not correctly accounted for. Second, given that linear corrections will induce
couplings of the form $(\ell_{(1,2)},\ell_{(2,1)}\pm1)$ and because of the restriction that
$\ell_1+\ell_2+\ell_3$ must be even in Eq.(\ref{kernel}), the effect uncovered is a
mixed-parity one. This means that $\beta$ corrections are identically zero for masks which
are even or odd functions. Real CMB masks do not possess
well-defined parity, a fact which further points to the importance of the analysis being
carried out here.

To illustrate the type of coupling induced by the matrices (\ref{matrixN}-\ref{matrixQ}),
we have carried out a numerical analysis using a north-south asymmetric equatorial strip as a
mask with different widths and degrees of asymmetry. Figure \ref{fig1} show some plots of
these matrices for $\ell_1=5$ as a function of $\ell_2$. Noticeably, the main difference between
(\ref{matrixN}-\ref{matrixQ}) and the matrix (\ref{matrixM}) is that the former assume
both positive and negative values, whereas the latter is a positive-definite matrix.
Regarding their amplitude, we have checked that matrices $\mathcal{N}$, $\mathcal{P}$ and
$\mathcal{Q}$ can be as large as 15\% of matrix $\mathcal{M}$ for the range of $\ell_2$
shown in Fig. \ref{fig1}. Including the $\beta$ factor, this should amount to a
correction of order $10^{-4}$ to Eq.(\ref{pseudo_Cls}). This may look smaller than the
known $10^{-3}$ signal present in the off-diagonal terms of full-sky correlation matrices 
\cite{Challinor:2002zh,Amendola:2010ty}. However, we emphasize that here we are 
proposing a correction to the reconstructed true $C_\ell$s, and not cross-terms of the
correlation matrix as discussed in \cite{Amendola:2010ty}; the former are not only less noise contaminated 
but also have a smaller 
cosmic variance.

%%%%%%%%%%%%%%%%%%%%%%%%%%%%%%%%%%%%%
\section{Discussion and Future Work}\label{discussion}
%%%%%%%%%%%%%%%%%%%%%%%%%%%%%%%%%%%%%

Determining our direction and velocity with respect to the CMB rest frame is a fundamental
quest for the standard cosmological model. While the main observable effects, noticeably the dipole, 
was already detected in the late 70's \cite{Smoot:1977bs}, higher multipolar distortions in the temperature
field may yet be uncovered by future high-resolution data from the Planck satellite. In this paper we have 
shown that the CMB power spectrum may be systematically contaminated by this effect already at
first order in the boost parameter if the boost effect is not taken into account when reconstructing full-sky data 
from cut-skies. We estimated this effect as a $10^{-4}$ signal in the pseudo $\mathcal{C}_\ell$s. Although this may seem 
smaller than the signal in off-diagonal full-sky CMB observables, only a careful reconstruction of full-sky data will 
be able to set this issue. We have also shown, using a simple strip as a mask, that the boost couples different 
multipoles in a way which is in sharp contrast to the couplings induced by the mask alone. Since the masking procedure is known to induce large-angle correlations,
reconstructing the $\mathcal{C}_{\ell}$s correctly may have a sizable impact in tracing systematics and/or
accounting for large-angle anomalies.

In a companion paper we will carry a more complete analysis of these effects, including the angle between the boost 
and the galactic plane as a free parameter. This angle dependence will lead to further couplings between the boost and 
the mask which might be used as a further estimator of the boost direction.

Since not all of the available information is contained in the $C_{\ell}$'s (due to the 
coupling between nearby modes of the $a_{\ell m}^{\prime}$), we will define off-diagonal
estimators of the covariance matrix. Additionally, we will determine the likelihood of
finding the direction of $\beta $ from our boosted covariance matrix. Both of these issues
have been discussed in part by \cite{Amendola:2010ty}. We will also be looking into the
effect of cut skies on reconstructing polarization estimators, which can be carried out starting from expression \ref{alm_freq_1}
 and setting $s=2$.

\appendix
%%%%%%%%%%%%%%%%%%%%%%%%%%%%%%%%%%%%%
\section{Second order results}
%%%%%%%%%%%%%%%%%%%%%%%%%%%%%%%%%%%%%

%%%%%%%%%%%%%%%%%%%%%%%%%%%%%%%%%%%%%
\subsection{Multipolar coefficients}
%%%%%%%%%%%%%%%%%%%%%%%%%%%%%%%%%%%%%
For completeness, we present here the expansion in the brightness multipolar coefficients to second
order in the boost parameter. Similar expansions can also be found in \cite{Challinor:2002zh,Amendola:2010ty}
We now want to expand Eq.~\ref{boost_alm} to $\mathcal{O}(\beta^{2})$. We begin by re-writing
\begin{equation}\label{expanded_gamma}
\gamma(1+\beta\cos\theta)\simeq 1+\beta\cos\theta+\frac{1}{2}\beta^{2}
\end{equation}
and $a_{\ell_{2} m_{2}}(\nu)$ using
\begin{equation}
a_{\ell m}(\nu)\simeq a_{\ell m}(\nu^{\prime})+
\left( \beta^{2}\cos^{2}\theta-\frac{\beta^{2}}{2}-\beta\cos\theta \right) \nu^{\prime}
\frac{d}{d\nu^{\prime}}a_{\ell m}(\nu^{\prime})+\left( \beta^{2}\cos^{2}\theta \right)
\frac{\nu^{\prime 2}}{2}\frac{d^{2}}{d\nu^{\prime 2}}a_{\ell m}(\nu^{\prime})
\end{equation}
We will also expand $_{s}Y_{\ell_{1} m_{1}}^{\ast}(\hat{\bf{n}}^{\prime})$ about 
$\hat{\bf{n}}$ to get
\begin{equation}
_{s}Y_{\ell m}(\hat{\bf{n}}^{\prime})=\, _{s}Y_{\ell m}^{\ast}(\hat{\bf{n}})-
\beta\sin\theta\left( 1+\beta\cos\theta \right)
{ \partial _{s}Y_{\ell m}^{\ast}(\hat{\bf{n}}) \over \partial
\cos\theta} \, 
+\frac{\beta^{2}\sin^{2}\theta}{2}{ \partial^{2}\, _{s}Y_{\ell m}^{\ast}(\hat{\bf{n}}) 
\over \partial \cos\theta^{2}}.
\end{equation}
There is a general expression for writing the derivatives of a spherical
harmonic in terms of other spherical harmonics (valid for $\ell\geq 1$)\cite{abramowitz+stegun}:
\begin{equation}\label{ylm_defn}
\sin\theta\,  {\partial _{s}Y_{\ell m} \over \partial \cos\theta}
=\ell\sqrt{\frac{(\ell+1)^2-m^2}{(2\ell+1)(2\ell+3)}}\, _{s}Y_{\ell+1,m}+ 
\frac{sm}{\ell(\ell+1)}\, _{s}Y_{\ell,m}
-(\ell+1)\sqrt{{\ell^2-m^2 \over (2\ell+1)(2\ell-1)}}\, _{s}Y_{\ell-1,m}
\end{equation}
We also see that there will be some $\beta \cos\theta$ cross terms to
deal with, which can be done using the formula
\begin{equation}\label{cosylm}
\cos\theta \, _{s}Y_{\ell m}
=\sqrt{{(\ell+1)^2-m^2 \over (2\ell+1)(2\ell+3)}}\, _{s}Y_{\ell+1,m}-
\frac{sm}{\ell(\ell+1)}\, _{s}Y_{\ell,m} +\sqrt{{\ell^2-m^2 \over (2\ell+1)(2\ell-1)}}\,
_{s}Y_{\ell-1,m} \; .
\end{equation}
Putting all this together and working to order $\beta^{2}$, we arrive at
an expression for the multipole moments measured in the moving frame
as a function of those in the CMB rest frame, noting that the usage of \ref{ylm_defn} and \ref{cosylm}
 limits this expression to $\ell\geq 2$:
\begin{eqnarray}\label{alm_freq}
a'_{\ell m}(\nu') & = & \left\{ 1-\frac{\beta
sm}{\ell(\ell+1)}\left(2-\nu'\frac{\text{d}}{\text{d}\nu'}\right)
+\beta^{2}\left[\frac{1}{2}{}_{s}\xi_{(\ell+1)m}^{2}\left(-\ell(\ell+4)+2\ell\nu'
\frac{\text{d}}{\text{d}\nu'}+\nu'^{2}\frac{\text{d}^{2}}{\text{d}\nu'^{2}}
\right)\right.\right.\notag \\
 &  &
+\frac{1}{2}{}_{s}\xi_{\ell
m}^{2}\left(-(\ell+1)(\ell-3)-2(\ell+1)\nu'\frac{\text{d}}{\text{d}\nu'}
+\nu'^{2}\frac{\text{d}^{2}}{\text{d}\nu'^{2}}\right)+\frac{1}{2}
\left(1-\nu'\frac{\text{d}}{\text{d}\nu'}\right)\notag \\
 &  &
\left.\left.+\frac{s^{2}m^{2}}{\ell^{2}(\ell+1)^{2}}\left(-\frac{3}{2}-\nu'\frac{\text{d}}
{\text{d}\nu'}+\frac{\nu'^{2}}{2}\frac{\text{d}^{2}}{\text{d}\nu'^{2}}\right)\right]
\right\} a_{\ell m}(\nu')\notag \\
 &  & -\beta{}_{s}\xi_{(\ell+1)m}\biggl[(\ell-1)+\nu'\frac{\text{d}}{\text{d}\nu'}-
\frac{\beta
sm}{\ell(\ell+2)}\left(2(\ell-1)-(\ell-2)\nu'\frac{\text{d}}{\text{d}\nu'}-\nu'^{2}
\frac{\text{d}^{2}} {\text{d}\nu'^{2}}\right)\biggr]a_{(\ell+1)m}(\nu')\notag \\
 &  & -\beta{}_{s}\xi_{\ell m}\left[-(\ell+2)+\nu'\frac{\text{d}}{\text{d}\nu'}-
\frac{\beta
sm}{(\ell+1)(\ell-1)}\left(-2(\ell+2)+(\ell+3)\nu'\frac{\text{d}}{\text{d}\nu'}-\nu'^{2}
\frac{\text{d} ^{2}}{\text{d}\nu'^{2}}\right)\right]a_{(\ell-1)m}(\nu')\notag \\
 &  &
+\frac{1}{2}\beta^{2}{}_{s}\xi_{(\ell+2)m}{}_{s}\xi_{(\ell+1)m}
\left(\ell(\ell-1)+2\ell\nu'\frac{\text{d}}
{\text{d}\nu'}+\nu'^{2}\frac{\text{d}^{2}}{\text{d}\nu'^{2}}\right)a_{(\ell+2)m}
(\nu')\notag \\
 &  &
+\frac{1}{2}\beta^{2}{}_{s}\xi_{\ell
m}{}_{s}\xi_{(\ell-1)m}\left((\ell+1)(\ell+2)-2(\ell+1)\nu'
\frac{\text{d}}{\text{d}\nu'}+\nu'^{2}\frac{\text{d}^{2}}{\text{d}\nu'^{2}}\right)
a_{(\ell-2)m}(\nu')
\end{eqnarray}
where 
\begin{equation}
_{s}\xi_{\ell m}=\sqrt{\frac{\left(\ell^{2} - m^{2}\right)\left(\ell^{2}-s^{2}\right)}
{\ell^{2}(2\ell + 1)(2\ell -1)}}.
\end{equation}
Note that this result differs slightly from the one shown in \cite{Challinor:2002zh}. We have checked that
they are the same up to  some re-arranging. 

%%%%%%%%%%%%%%%%%%%%%%%%%%%%%%%%%%%%%
\subsection{Expansion of the Pseudo Angular Power Spectrum}
%%%%%%%%%%%%%%%%%%%%%%%%%%%%%%%%%%%%%
Using expression (\ref{alm_freq}) and (\ref{pseudo_aniso}) we can show after a straightforward algebra that:
\begin{widetext}
\begin{eqnarray}\label{boostedpseudo}
%\lefteqn{}
\langle \mathcal{\tilde{C}}_{\ell_{1}}^{\prime}\rangle & = &
\sum_{\ell_{2}}C_{\ell_{2}} M_{\ell_{1}\ell_{2}}+\frac{1}{(2\ell_{1}+1)}
\sum_{\ell_{2}}\sum_{m_{1}m_{2}}K_{\ell_{1}m_{1}\ell_{2}m_{2}} \Big\{  \beta \big [
\xi_{\ell_{2}+1m_{2}}(\ell_{2}+4)K_{\ell_{1}m_{1}\ell_{2}+1m_{2}} \notag\\
&&
 -\xi_{\ell_{2}m_{2}}(\ell_{2}-3)K_{\ell_{1}m_{1}\ell_{2}-1m_{2}}\big ] C_{\ell_{2}}  + \big [ -\beta
\xi_{\ell_{2}+1m_{2}}(\ell_{2}-2)K_{\ell_{1}m_{1}\ell_{2}+1} \big ] C_{\ell_{2}+1} \notag\\
&& + \big [ \beta
\xi_{\ell_{2}m_{2}}(\ell_{2}+3)K_{\ell_{1}m_{1}\ell_{2}-1m_{2}}\big ] C_{\ell_{2}-1}\Big\}.
\end{eqnarray}
\end{widetext}

\begin{acknowledgments}
We would like to thank Anthony Challinor, Craig Copi, Arthur Kosowsky, Dominik J. Schwarz, and Pascal Vaudrevange for useful conversations 
during the preparation of this work. TSP thanks Brazilian agency FAPESP for partial
support and the physics department of Case Western Reserve University for its hospitality
during the initial stages of this work. GDS and AY are supported by a grant from the US Department of Energy and by 
NASA under cooperative agreement NNX07AG89G. MS is supported by the Friedrich- Ebert-Foundation and thanks  
the physics department of Case Western Reserve University for its hospitality.

\end{acknowledgments}

%%%%%%%%%%%%%%%%%%%%%%%%%%%%%%%%%%%%%
%\section{References}
%%%%%%%%%%%%%%%%%%%%%%%%%%%%%%%%%%%%%

\bibliographystyle{h-physrev}
\addcontentsline{toc}{section}{\refname}\bibliography{CMB_boost}

\end{document}